# Erosion rate of lunar soil under a landing rocket, part 2: benchmarking and predictions

**Authors:**

Philip T. Metzger[1]*

**Affiliations:**

[1]Stephen W. Hawking Center for Microgravity Research and Education, University of Central Florida, Orlando, FL 32828, USA.

*Corresponding author. Email: Philip.metzger@ucf.edu

Abstract: In the companion paper ("Erosion rate of lunar soil under a landing rocket, part 1: identifying the rate-limiting physics", *this issue*) an equation was developed for the rate that lunar soil erodes under the exhaust of a landing rocket. That equation has only one parameter that is not calibrated from first principles, so here it is calibrated by the blowing soil's optical density curve during an Apollo landing. An excellent fit is obtained, helping validate the equation. However, when extrapolating the erosion rate all the way to touchdown on the lunar surface, a soil model is needed to handle the increased resistance to erosion as the deeper, more compacted soil is exposed. Relying on models derived from Apollo measurements and from Lunar Reconnaissance Orbiter (LRO) Diviner thermal inertia measurements, only one additional soil parameter is unknown: the scale of increasing cohesive energy with soil compaction. Treating this as an additional fitting parameter results in some degeneracy in the solutions, but the depth of erosion scour in the post-landing imagery provides an additional constraint on the solution. The results show that about 4 to 10 times more soil was blown in each Apollo landing than previously believed, so the potential for sandblasting damage is worse than prior estimates. This also shows that, with further development, instruments to measure the soil erosion during lunar landings can constrain the soil column's density profile complementary to the thermal inertia measurements, providing insight into the landing site's geology.

1. <u>Introduction</u>

Erosion of soil under a landing rocket is different than terrestrial soil erosion and dust emission. The latter are both dominated by saltation transporting energy from the gas down to the granular surface, but rocket exhaust is so fast that saltation does not occur. Ejected particles do not re-impact the surface for hundreds to thousands of kilometers, if at all (fines can be ejected completely off the Moon), and the gas expands into vacuum, so erosion is limited to a radius around the rocket much smaller than the distance at which particles re-impact. Leading up to the Apollo Moon landings, Roberts [1] hypothesized that erosion under these conditions should be linear with the shear stress of the gas, thinking that erosion would be a shearing process and that shear strength of the soil would resist the erosion. He hypothesized that the erosion rate would be mediated by the acceleration of particles already liberated from the surface, which removes momentum from the gas until its remaining shear stress no longer exceeds the shear strength of the soil. This hypothesis has been used in recent studies to prepare for lunar landings [2,3], so it is important to know if it is correct.

The companion paper ("Erosion rate of lunar soil under a landing rocket: identifying the rate-limiting physics", *this issue*) argues that Roberts' erosion hypothesis is incorrect. As Metzger et al. [4] earlier argued, the acceleration of particles continues far downstream of the erosion point, so (in supersonic flow, especially) it is impossible for them to govern the erosion rate that develops upstream. Also, erosion should not be restrained by the shear strength of the soil since particles are lifted away from the free surface rather than being sheared horizontally across it. Furthermore, dimensional analysis of the experimental results show that the erosion rate should be related to energy flux of the gas rather than its shear stress.

The companion paper also shows that, without saltation, the erosion of soil by a rocket exhaust is a surprisingly low-energy process. The eroded particles do achieve high energy after being lifted from the surface and exposed to the higher velocity gas above the boundary layer, so they are accelerated to high speed and can cause unacceptable sandblasting damage to other assets on and around the Moon [5–7]. However, the energy consumed accelerating the grains downwind has negligible effect on the erosion process that occurs below the laminar sublayer, so it does not affect the erosion rate equation. It is the tiny fraction of kinetic energy that molecularly diffuses to the bottom of the laminar sublayer that participates in lifting particles from the surface. Resistance to erosion, and hence erosion rate, is therefore governed by two energies that are properties of the soil: the potential energy to lift grains to the tiny height on the order of a grain diameter, and the cohesive energy of the soil.

With this new theory, there is only one remaining unknown parameter to calibrate: the erosion efficiency, which is the fraction of this energy at the bottom of the laminar sublayer that is converted into mechanical work lifting the grains. To calibrate it, section 2 of this paper develops the equations of dust opacity in lunar landings. Section 3 uses these equations to predict the optical density of the blowing soil during the Apollo 16 landing, and these predictions are used to test the theory and to calibrate the erosion efficiency. With this result, it is possible to integrate the erosion at each location on the soil during descent of the rocket, determining depth of scour. This shows the loose dust and soil is removed to a sufficient depth that the properties of the deeper, more compacted soil become relevant. Therefore, Section 4 develops a model of cohesive energy density of the soil as a function of its depth and compaction. Section 5 integrates the erosion equation with this soil model through the entire landing trajectory, determining how the soil model is constrained by the existing data sets. This determines the plausible range of ejected soil mass, finding it to be much higher than previously believed. Section 6 discusses limitations of this study and the need for additional research including data taken during actual landings on the Moon.

## 2. **Theoretical Prediction of Dust Opacity**

*2.1. Erosion Rate Equation*

As shown in the companion paper, the height $\langle D \rangle$ above the surface is the height that an individual soil grain must be lifted to be swept downwind by the gas, approximately 1.5 $D_{84}$, or about one coarse grain diameter. ($D_{84}$ is the diameter such that 84% of the mass of the soil is finer. It is a commonly used diameter in sedimentary geology.) The downward energy flux from the gas crossing this height $\langle D \rangle$ is $E = \rho_0 v_0^2 \, \overline{v_T}/12$, where $\rho_0$ and $v_0$ are the local gas density and velocity flowing horizontally at or just above $\langle D \rangle$ and $\overline{v_T}$ = the mean thermal velocity of the gas molecules. The equation developed in the companion paper is

$$\dot{m} = \rho_b \frac{\varepsilon \, (E - E_{\text{th}})}{\rho_b \, g \, \langle D \rangle + \alpha}, \qquad E > E_{\text{th}} \tag{1}$$

or $\dot{m} = 0$ if $E < E_{\text{th}}$, where $E_{\text{th}}$ = threshold value of $E$ capable of initiating erosion, $\dot{m}$ = local mass erosion rate (kg/m²/s), $\rho_b$ = bulk density of the soil (kg/m³), $\varepsilon$ = efficiency (unitless) of the energy flux crossing the plane at height $\langle D \rangle$, $g$ = gravitational acceleration (m/s²), and $\alpha$ = cohesive energy density in the soil (J/m³). All the parameters can be determined or at least estimated from first principles except for $\varepsilon$, which is not well enough understood.

$\varepsilon$ can be calibrated empirically via the opacity of blowing dust in lunar landings. Apollo 16 is chosen for this because the ejecta cloud contains fewer discrete streaks than the other Apollo landings (caused by larger craters and rocks in the erosion zone locally enhancing shear stress) so it is better for averaging the brightness of the dust in the cloud.

*2.2. Dust Opacity Equations*

Brightness of an image pixel in the LM descent video is modeled following Lane and Metzger [8] and You et al. [9] as

$$B = B_1(1 - e^{-X}) + B_2 e^{-X} \tag{2}$$

The first term is the contribution of multiple scattering in the cloud. The second is the contribution of the direct path from the Sun to the lunar surface then up to the camera. $X$ = optical depth of the dust, which is found by integrating the light scattering along the path of the sunlight through the cloud. $B_1$ and $B_2$ are collections of parameters that include the camera's white and black levels, their conversion to digital video, and the soil's albedo. They cannot be calibrated directly because it was a film camera, and the film from fifty years ago cannot be reconstituted to permit testing. However, $B_2 = 0.477$ was determined from the videos by averaging the brightness of the ground before the dust begins blowing. $B_1$ will be a fitting parameter along with $\varepsilon$.

Soil particles leave the surface in a flat ejecta cone with ~3° elevation above the local horizontal and fills the space below that cone, which was measured via the shadows of the Lunar Module (LM) [10] and confirmed in computer simulations [11].

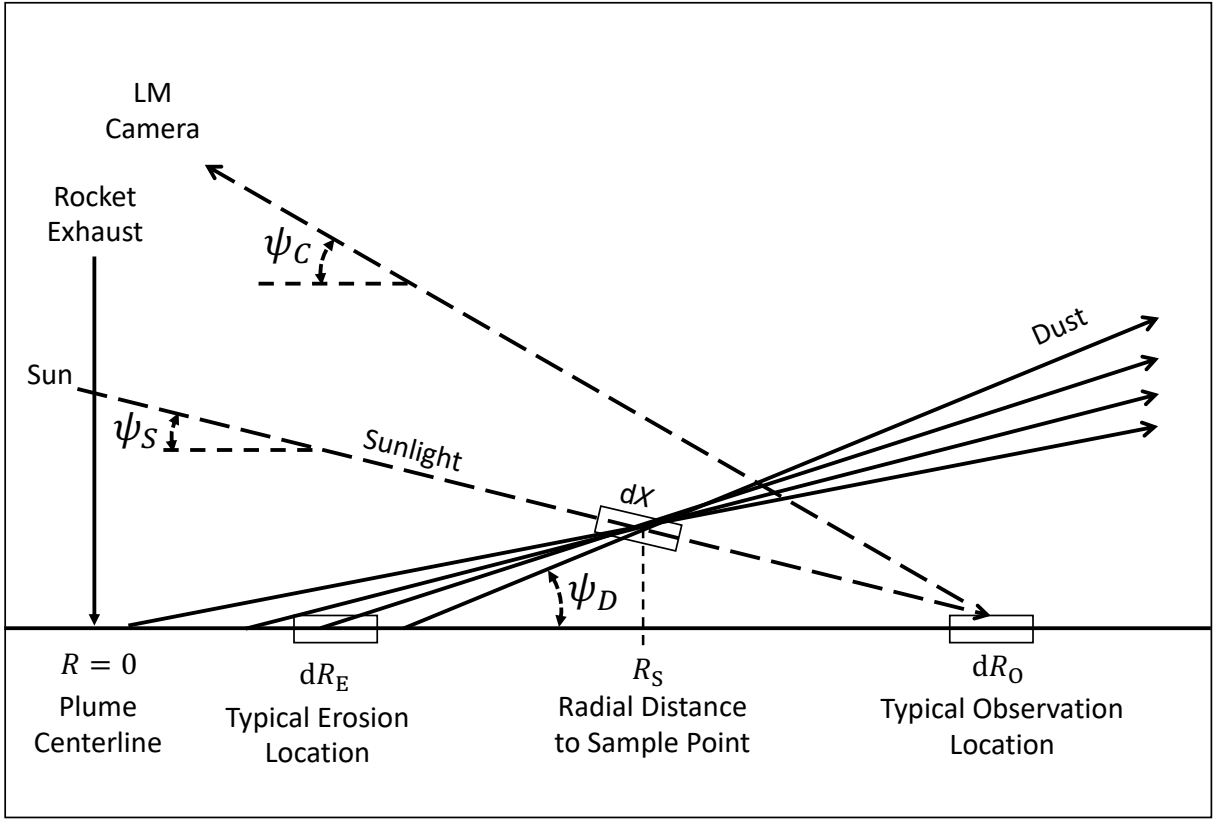

**Fig. 1. Geometry of dust erosion, sunlight, and camera.**

Integrating the light along the path shown in fig. 1,

$$X = \int_{0.013\,\mu m}^{1000\,\mu m} dD\, \frac{\pi D^2 P(D)}{4\, v(D)} \int_{380\,nm}^{750\,nm} d\lambda\, \frac{S(\lambda)}{(750\,nm - 380\,nm)} C_E(D,\lambda)$$

$$\times \int_{\xi R_0}^{R_0} dR_S \left\{ \int_0^{\hat{R}(\psi_C)} dR_E\, \frac{\dot{m}(R_E, H, \widehat{D}) R_E \sec(\psi_C)}{R_S(R_S - R_E)\tan(\psi_D)} + \int_0^{\hat{R}(\psi_S)} dR_E\, \frac{\dot{m}(R_E, H, \widehat{D}) R_E \sec(\psi_D)}{R_S(R_S - R_E)\tan(\psi_D)} \right\}$$

(3)

where $C_E$ = the single particle coefficient of extinction in Mie and Rayleigh scattering, $S(\lambda)$ = the solar spectrum, and $P(D)$ = the particle size distribution of lunar soil. The lower limit of integration for soil particle sizes is the smallest size detected in lunar soil by Park et al.[12]. The upper limit was chosen large enough that the integral is insensitive to making it larger. The limits of integration on the wavelengths represent the visible range, assuming the Apollo camera film was sensitive to that range. Computational fluid dynamics simulations have shown that finest particles blow generally faster than coarse ones in a rocket plume, so they are "stretched out" through space much less densely [11,13,14]. The entrained particle size distribution becomes $P(D)/v(D)$, where $v(D)$ is the average velocity of each size [8]. $H$ = height of the lander, $R_0$ = radial distance to the point on the regolith where the camera is aimed, $R_E$ = radial distance to a point on the regolith where erosion is occurring. $R_S$ = radial distance to the sample point, which is a differential portion of the path length (either down or up) where light is scattering in the dust. $\xi = 1/(1 + \tan(\psi_D))/\tan(\psi_S)$ is multiplied onto $R_0$ to find the nearest radial distance at which the downward path of sunlight toward $R_0$ intersects the dust cloud. $\psi_D = 3°$ is the upper bounding angle of the eroded dust trajectories from each erosion point. $\psi_S = 11.9°$ was the Sun angle for this mission. $\psi_C = 57°$ is the look angle of the camera determined by Lane and Metzger [8]. The last two integrals that are summed are for the upward and downward paths of the light. They differ only in their upper limits of integration,

$$\hat{R}(\psi_C) = R_0 - (R_0 - R_S)\frac{\tan(\psi_C)}{\tan(\psi_D)}, \quad \hat{R}(\psi_S) = R_0 - (R_0 - R_S)\frac{\tan(\psi_S)}{\tan(\psi_D)} \quad (4)$$

which are the furthest points from the plume centerline from which dust can erode and then intersect the measured light path(s).

*2.3. Particle Size Distribution*

Integrating eq. 3 requires the particle size distribution $P(D)$. Most measurements of $P(D)$ for actual lunar soil are reliable only down to 44 microns size [15]. Almost none report sizes below 2 or 10 microns. Greenberg et al. [16] and Park et al. [12] measured the ultrafine range. The results of Greenberg et al. are bimodal with deep dips at 0.5 microns, probably a measurement artifact [12,17]. Park et al. avoided the artifact but used samples that they dry sieved for <43 $\mu$m without reporting the relative masses of the splits, and there were no particles counted larger than 12 $\mu$m in the analyzed portion of the fines split, so merging the fines fraction with the coarse fraction is ambiguous. Here, the ultra-fines of Apollo sample 10084 measured by Park et al. were merged with a curve of the coarser fraction of lunar simulant JSC-1A (because it represents "average" lunar soil across the coarser range) setting the fines fraction <10 $\mu$m to 6%wt for the baseline case. This 6%wt was estimated from the Lunar Sourcebook's figure 9.1 [18]. Due to its uncertainty, the value shall be varied over a large range, below. Figures 2 and 3 use this merged $P(D)$ to demonstrate the dominant role of the ultra-fines fraction in both optical density of the blowing dust and in cohesion in the soil resisting erosion.

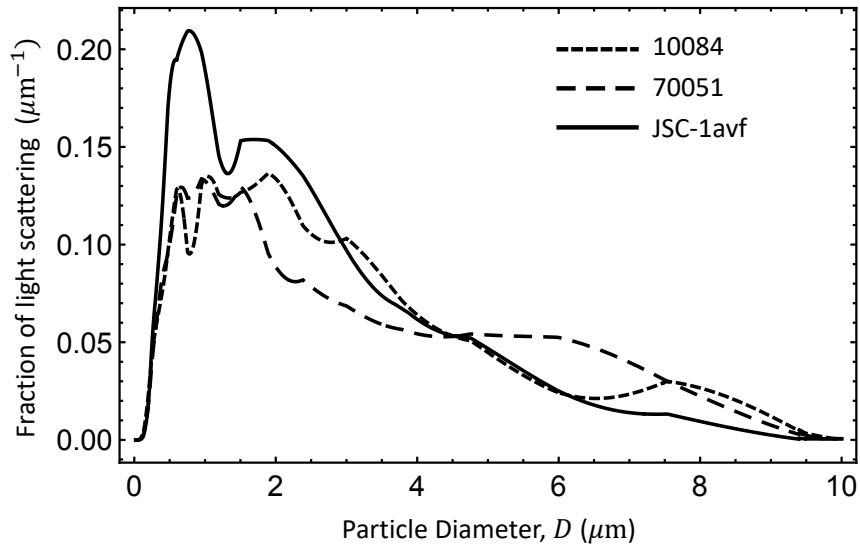

**Figure 2. Fraction of total light scattering per micron of particle size.** This was calculated by multiplying the cross sectional area of a particle of size $D$ with the extinction coefficient for that size and $P(D)$, divided by the integral of the same over all particle sizes. It was performed for each of the three fines samples of Park et al. [12] at 6%wt fines fraction.

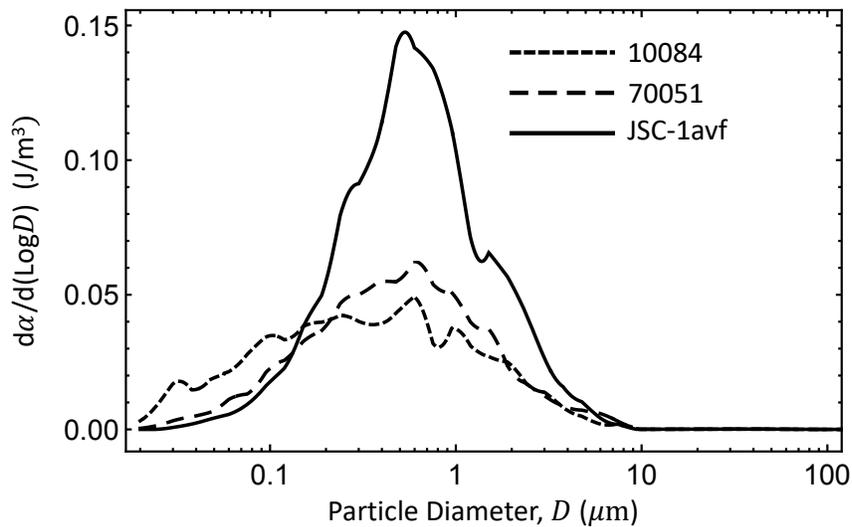

**Figure 3. Cohesive energy density $\alpha$ per log particle size.** This was calculated with the van der Waals equation for a pair of particle sizes in contact, integrating it over the distribution of sizes for one of the contacting particles, then multiplying by the distribution of sizes for the other contacting particle. It was performed for each of the three fines samples of Park et al. [12] at 6%wt fines fraction.

*2.4. Gas Flow Conditions*

Roberts' gas flow equations (but not his soil erosion equation, which is incorrect) are used to calculate $\rho_0$, $v_0$, $\overline{v_T}$, and laminar $\tau$ as functions of vehicle altitude and radial distance from centerline. This enables calculation of $\dot{m}$ via eq. 1, which is needed for eq. 3. The flow is generally laminar since rarefaction in lunar vacuum suppresses turbulence. The parameters in Roberts' equations were set for a lander with mass 7.5 t representing the J-missions (Apollo 15-17) at touchdown, with 2 crewmembers, equipment, and consumables, with aerozine/nitrogen tetroxide propellant and a nozzle having the expansion ratio and exit diameter of the J-missions. The combustion chamber pressure was set for a thrust-to-weight ratio $= 1$ in lunar gravity for the observed constant rate of descent. Dust blowing is first observed in the Apollo 16 landing video when the LM is at 31.5 m altitude, so the energy flux at radius $R$ that is the maximum value of $E(R)$ for this altitude is the threshold. It is found to be $E_{\text{th}} = 0.123$ J/m²/s.

## 3. Comparing Empirical Data to Theory

### 3.1. Available Data

The Apollo 16 landing video for comparison with this theory was from the NASA History Office's Apollo Flight Journal [19,20]. Two video frames are compared in fig. 4, showing that the histogram of pixel brightness becomes narrower and shifts to the right as the dust is blowing. Lane and Metzger [8] used a "Histogram Matching Method" to quantify the dust in each frame, but here it is adequate, simpler, and possibly more accurate to use the mean brightness instead.

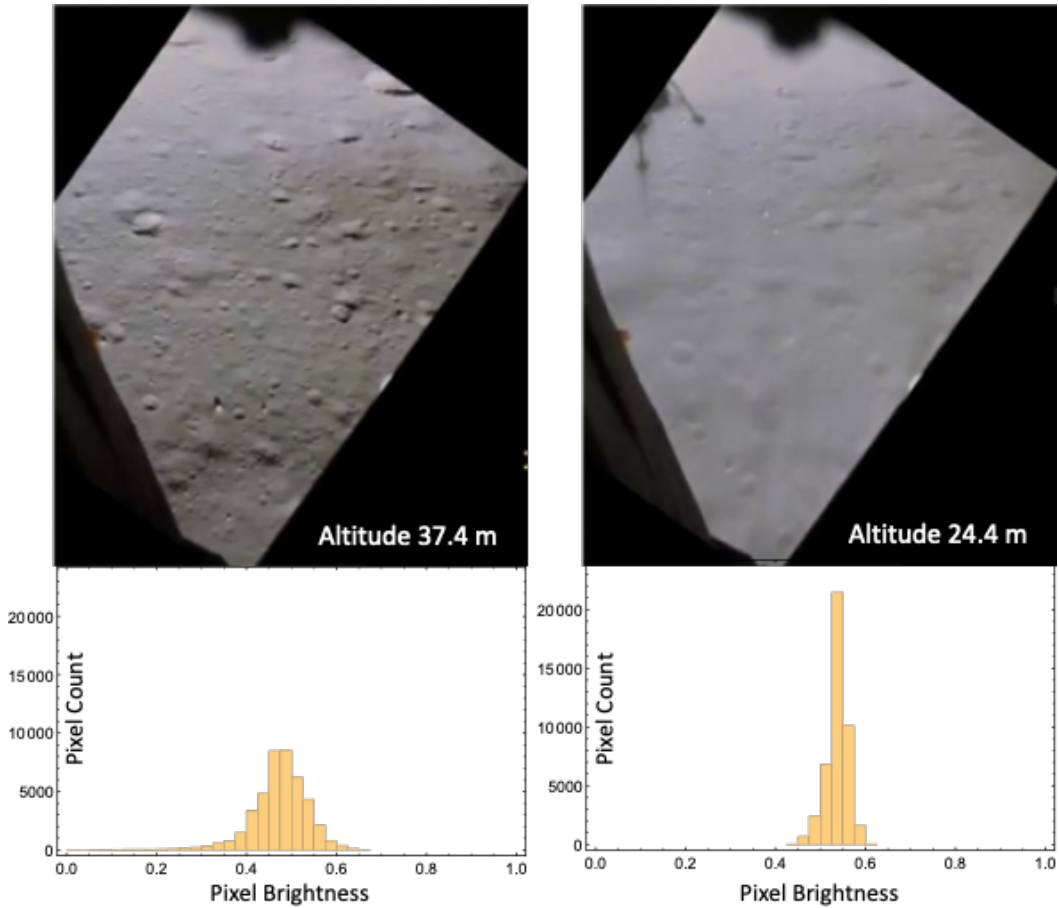

**Figure 4. Changes in pixel brightness histogram due to soil erosion.** Comparison for video frames at 37.4 m and 24.4 m altitude, before and after dust blowing began. (Video frames credit: Apollo Flight Journal/NASA. Used by permission.)

*3.2. Fitting Theory to the Data*

$X$ was calculated via eq. 3 for altitudes between 32 and 15 m, which is the usable range of the video before the LM shadow interferes. Future work could include lower altitudes by carefully cropping around the moving shadow. A guessed value of $\varepsilon$ was initially used, but eq. 3 is proportional to the constant $\varepsilon$ via $\dot{m}$ in the integrand, so $X$ can be scaled up or down linearly for different values of $\varepsilon$ to best fit to the videographic data rather than reperforming the integrations in each iteration. (This is true only for homogenous soil models, so in section 5 where erosion rate depends on the depth of prior erosion the full model must be reperformed each time.) Integrating one landing trajectory takes several hours on a laptop computer running Mathematica.

Using the baseline size distribution $P(D)$ for lunar soil, the best fit to the Apollo 16 landing data is with $\varepsilon = 0.0029$ and $B_1 = 0.571$, shown in fig. 5. When $\varepsilon$ is 20% larger or smaller, then no value for $B_1$ can fit the opacity curve, confirming that the results are not overfitted. This value of $\varepsilon$ indicates 0.29% of the energy that diffuses below $\langle D \rangle$ converts to mechanical erosion while the rest constantly diffuses back up above $\langle D \rangle$ and is advected away in the lateral gas flow. This is plausibly the instantaneous mass fraction of the gas below $\langle D \rangle$ that is beneath erodible grains, so $\varepsilon$ might be purely geometric, explaining its lack of strong dependence on the other parameters. In future work it might be estimable from first principles.

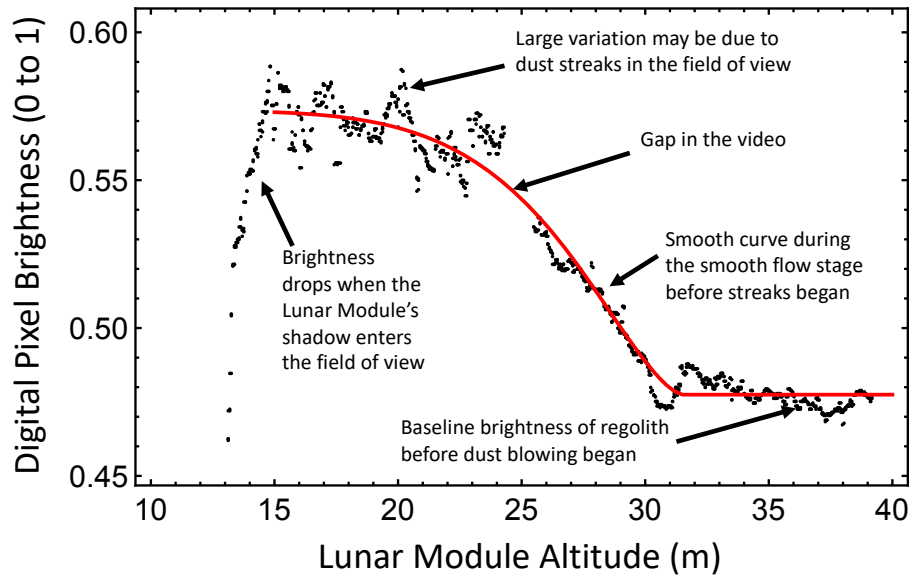

**Figure 5. Validating the theory via dust opacity in Apollo 16 landing.** Black points: average pixel brightness in a central box of the camera's field of view. Smooth red curve: fitted theory.

## 4. Lunar Soil Column Model

With a calibration of $\varepsilon$, eq. 1 enables extrapolation of the erosion rate to lower altitudes where the dense cloud prevents visible light cameras from measuring $X$. However, the erosion initially removes the looser surface soil, exposing denser soil that is more resistant to erosion, slowing the rate. This slowing is so severe that when it is neglected the extrapolation predicts 169 tons of soil were blown creating a 6.5 m deep crater, which is obviously incorrect. What has been omitted is a realistic model of the soil column.

*4.1. Available Data*

The strata in the column are impact ejecta blankets of varying thickness [21] that lay on the surface subject to micrometeoroid gardening for varying periods before burial by additional ejecta blankets. Unless we have data on the specific soil column of the landing site, we will be unable to predict exactly how the erosion rate varies with depth. By using functions that represent the lunar average, we can derive an average expectation. Carrier et al. [18] compiled the core tube and drill core data from multiple locations near the Apollo 15-17 landing sites and found the average trend,

$$\rho_b(z) = 1920 \frac{z + 0.122}{z + 0.18} \ (\text{kg/m}^3)$$

where $z$ = depth in m into the soil column, $\rho_0 = \rho_b(0) = 1300$ kg/m³ (58.1% porosity), and $\rho_\infty = \rho_b(z \to \infty) = 1920$ kg/m³ (38.0% porosity).

The shallowest datapoint that informed this curve was $z \cong 19$ cm, well below the depth of interest for plume erosion. The best available data to attempt constraining $\rho_b(z)$ more accurately in the relevant range, $0 < z < 10$ cm, are from astronaut boot prints, vehicle tracks, and remote sensing. The remote sensing data include optical scattering and solar wind scattering, which probe $z \approx 1$ mm, and thermal inertia that probes to $z \approx 10$ cm (insolation dependent). The optical density of blowing dust during landing is shown here to be another data set that constrains the soil column density to $z \approx 10$ cm.

Namiq [22] calibrated the boot prints via lunar simulant tested in Earth's gravity coupled with finite element modeling. Houston et al. [23] analyzed 323 boot prints from Apollo 11, 12, 14, and 15 using Namiq's calibration and reported an average surface porosity of 43.3% for intercrater areas, corresponding to an average penetration depth of only $z = 0.61$ cm and bulk density 1760 kg/m³. Tracks from the Apollo program's Lunar Roving Vehicle (LRV) were $z = 0$ to 5 cm deep, averaging 1.25 cm [24]. The Modular Equipment Transporter (MET) and Lunakhod rovers had similar wheel ground pressure, and analysis of all three sets of tracks concluded that the softer regions of soil had 47% porosity (1640 kg/m³) while firmer soils had 39% to 43% porosity (1770 to 1890 kg/m³). These porosities seem too low compared to the other datasets, below, perhaps due to the calibration, perhaps also skewed statistically by many boot prints and tracks in the area where the lander plume had blown away the looser layers.

To explain the surficial optical properties, Mendell and Noble [25] hypothesized a ubiquitous surficial layer or epiregolith ~90% porous at least 250 µm or 5-10 grain diameters thick. Ohtake et al. [26] used visible to near infrared reflectance at the collection site of Apollo sample 62231 measured by the SELENE spacecraft comparing to laboratory measurements of the returned sample, finding they agreed if the sample had porosity 74% to 86%. Hapke and Sato [27] reworked the analysis for the same site to find $83\pm3\%$ porosity, or possibly higher if certain assumptions were wrong. Szabo et al. [28] modeled solar wind protons scattering from the lunar surface as neutral hydrogen and compared it to measurements from Chandrayaan-1, which found a global average porosity of $85\%^{+15\%}_{-14\%}$ to a depth of at least 5 grain diameters. To be compatible with the wheel tracks and boot prints (regardless of their calibration), this hyperporous epiregolith must be very thin with more highly compacted soil immediately beneath it. Also, the modeling of eroded dust, below, could not match the observed brightness curve unless the epiregolith was assumed to be negligible in thickness so it blew away immediately having no effect on the optical density. When it was assumed to have non-negligible thickness, the value of $\varepsilon$ would be calibrated by the loose soil that began blowing first, then when more compacted soil with higher erosion resistance was exposed the erosion rate slowed too quickly such that no combination of the remaining parameters could match the brightness curve. Together, these observations indicate that the epiregolith is real but extremely thin and may be ignored in the erosion analysis.

To measure thermal inertia, Hayne et al. [29] analyzed the lunar night cooling curve of the regolith measured by Lunar Reconnaissance Orbiter (LRO) Diviner. They assumed an exponential decay for the soil density with depth,

$$\rho_b(z) = \rho_\infty - (\rho_\infty - \rho_0)(1 - e^{-z/H})$$

(5)

To avoid model degeneracy, they assumed $\rho_0 = 1100$ kg/m³ (41.9% porosity) and $\rho_\infty = 1800$ kg/m³ (64.5% porosity) for all lunar locations. The best fit varied by location from $H = 0$ cm to >15 cm with a mean value $H = 6.8$ cm. For comparison, integrating eq. 5 with $H = 6.8$ cm over the mean track depth of 1.5 cm depth finds an average porosity 62% (1170 kg/m³), much looser than the analysis of wheel sinkage indicated.

The Lunar Sourcebook soil model broadly agrees with the thermal inertia model and is similarly inconsistent with the boot print and track calculations. The Lunar Sourcebook model for 19 cm depth, where it is anchored by data, is 1619 kg/m³, close to the value from the boot print and track analysis at the surface. Therefore, if the boot print and track analysis is accurate, the top 19 cm would have to be roughly constant density. This shall be one of the models tested below with the erosion brightness curve. Another model, assuming the boot print and track analysis is incorrect, will be the thermal inertia model of eq. 5 but with surface densities set to either $\rho_0 = 1100$ kg/m³ or $\rho_0 = 1300$ kg/m³. When using 1300 kg/m³, the $H$ value shall be adjusted ad hoc to 8 cm to attempt compensating for the lower thermal inertia of the upper surface and to test the sensitivity of the soil column parameters.

*4.2. Cohesive Energy Density*

The cohesive energy at the surface of the soil, $\alpha_0$, is estimated per the van der Waals equation [30],

$$\alpha_0 = Z\, N_{V,0} \iint_0^\infty dD_1\, dD_2\, 2\pi\, e_s \left(\frac{D_1 D_2}{D_1 + D_2}\right)\left(h_{\text{asp}} - \frac{h_{\text{asp}}^2}{z_{\text{max}}}\right) P(D_1) P(D_2)$$

(7)

where $Z$ = average number of grain-to-grain contacts per grain, $D_1$ and $D_2$ are the diameters of two contacting grains, $e_s$ = the surface energy of the contact, $h_{\text{asp}}$ represents the height of asperities on the surfaces that prevent the grains from perfectly touching, and $z_{\text{max}}$ = the separation distance between grains beyond which there is no force of attraction. Per Götzinger and Peukert [31] and Israelachvili [32], $e_s = 0.05$ J/m², $h_{\text{asp}} = 1.65 \times 10^{-10}$ m, and $z_{\text{max}} = 1$ µm. $N_{V,0} = \chi_0/\bar{V}_{\text{gr}}$ is the number of grains per volume at the surface, where $\chi_0 = \rho_0/\rho_m$ is the compactivity of soil at the surface, $\rho_m = 3100$ kg/m³ is the typical mineral density of lunar soil, and $\bar{V}_{\text{gr}}$ is the average volume per grain. The latter is calculated by integrating the volume of a sphere over $P(D)$. $Z \approx 3$ is assumed, each grain having ~6 grain-to-grain contacts but each contact shared by 2 grains. For example, using $\rho_0 = 1000$ kg/m³ and using the baseline $P(D)$ finds $\alpha_0 = 0.289$ J/m³.

Lunar soil is hyperstatic so $Z$ increases with bulk density, notably as ultra-fines form necks with many contacts between the larger grains as illustrated in fig. 6. This causes $\alpha$ to increase with depth in the soil column where the soil is denser. Nguyen et al. [33] modeled polydisperse granular packings and found that $Z$ increases more for the larger particles than the smaller ones as the span of the polydispersity increases, and this is suggestive of bridging or necking between the larger grains, but modeling lunar soil where polydispersity spans five orders of magnitude is computationally too expensive. An empirical method is needed to quantify $\alpha$ as it increases with compaction.

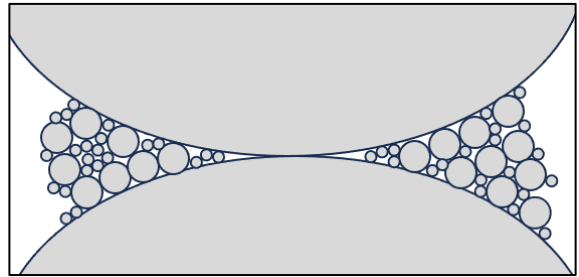

**Figure 6. Illustration of finer grains forming necks between larger grains.** This increases the effective surface area of the contacts between larger grains, holding the fabric of the soil together with more cohesive energy and increasing thermal conductivity through the soil.

Another property of lunar soil controlled by necking of the finest grains between the coarser grains is thermal conductivity. Metzger et al. [34] showed that the increase in thermal conductivity with bulk density is consistent with an exponential form, so by analogy cohesion ought also to follow an exponential. Direct shear measurements indeed discover an exponential form for the increase in Mohr-Coulomb cohesion with bulk density [35]. This cohesion from the Mohr-Coulomb relationship includes both van der Waals force and "apparent cohesion" from the shear resistance of interlocking grains [36], so it is related to, but not identical to, the cohesion that resists pulling grains perpendicularly from a free surface.

These considerations indicate that the exponential form is probably correct for the van der Waals cohesion and it shall be used for the model here,

$$\alpha(z) = \alpha_0 \, \text{Exp}\left[\frac{\rho_b(z) - \rho_0}{k}\right] \tag{6}$$

where $k$ is a parameter that shall be treated as a fitting parameter in the next section. Future laboratory work with lunar samples should attempt to measure $k$ directly.

The surface conditions $\alpha_0$ and $\rho_0$ determine the initial values of both erosion threshold $E_{\text{th},0}$ and erosion rate $\dot{m}_0$. Where deeper soil has been exposed the scaling is,

$$E_{\text{th}}(z) = E_{\text{th},0} \left( \frac{\rho_0 \, g \, \langle D \rangle + \alpha_0}{\rho_b(z) \, g \, \langle D \rangle + \alpha(z)} \right) \tag{7}$$

$$\dot{m}(z) = \dot{m}_0 \left( \frac{\rho_0 \, g \, \langle D \rangle + \alpha_0}{\rho_b(z) \, g \, \langle D \rangle + \alpha(z)} \right) \tag{8}$$

where the baseline $P(D)$ from section 2.3 is used to calculate $\langle D \rangle = 1.5 \, D_{84}$.

## 5. **Integration Over Entire Landing Trajectory**

The procedure is to integrate eq. 3 using eq. 1 with a soil model for $\rho(z)$ and $\alpha(z)$, iterating $\varepsilon$, $B_1$, and the soil model parameters $k$ for consistency with the empirical data. More free parameters would be under-constrained, so the existing soil models are used for $\rho_0$, $\rho_\infty$, and $H$. It was found that $\varepsilon$ primarily controls the slope of the brightness curve of fig. 5 when the lander is at high altitudes, $B_1$ primarily controls the amplitude of the brightness curve after it levels off (saturation of the exponentials in eq. 2), and $k$ primarily controls the depth of the crater after landing. All three parameters have some (smaller) effect on all the observables, so iteration can fine-tune the solution.

The depth of the crater after landing was not directly measurable in the Apollo post-landing imagery because the crater is shallower and more gradual than the natural terrain variations, and because it is spread over an area wider than the images. However, Metzger et al. [4] used thruster plume tests on volcanic tephra at a field site to interpret the erosional features, which included both headed and unheaded erosional remnants and micro-scarps. From this they estimated about 6 cm were eroded at the deepest point between 1 and 2 m radius from centerline in a toroidal crater. The extremes of plausibility for this estimate are 3 cm to 12 cm depth. The procedure here is to pick a depth within the plausible range and find the corresponding solution for $\varepsilon$, $B_1$, and $k$. This is repeated for several depths to obtain the range of possible parameter values. $\varepsilon$ and $B_1$ were found to vary only weakly across this range but $k$ varies greatly. Therefore, most of the uncertainty in this method is from $k$.

Soil Model 1 assumes the boot print and rover track data are calibrated correctly with constant density $\rho_0 = 1600$ kg/m³ in the top 19 cm so it does not need a value for $k$. Fitting to the brightness curve found $\varepsilon = 0.003$ and $B_1 = 0.579$. This resulted in a crater that was 5.15 m deep at 82 cm radial distance from centerline, which is clearly wrong per the post-landing imagery. This shows that increasing resistance with depth is crucial to the physics, falsifying the idea that the bulk density might be constant in the top of the soil column to the depth eroded by the plume. Trusting the core tube and drill core data are correct we may conclude that the boot print and wheel track data did indeed have a calibration bias.

Soil Model 2 follows Hayne et al. [29] with $\rho_0 = 1100$ kg/m³ and the lunar average $H = 6.8$ cm. Excellent fits to the brightness curve were obtained for craters 5 cm and deeper, but 4 cm and smaller craters did not permit a good fit. There is a modest range of degeneracy in the selection of parameters, and $\varepsilon$ was selected to be the same in each of these cases. The results are given in Table 1.

Soil Model 3 also used the functional form of Hayne et al. but followed Carrier et al. with $\rho_0 = 1300$ kg/m³, and it used $H = 8$ cm in part to compensate (in approximation) for the higher thermal inertia of $\rho_0$ but mainly to simply try another value. As before, excellent fits to the brightness curve were obtained for craters 4 cm and deeper, but a 3 cm crater did not permit a good fit. For this soil model, $\varepsilon$ was iterated for optimal fit rather than kept constant between cases, to test parameter sensitivities a different way. The results of each soil model and choice of erosion depth are given in Table 1. Figs. 6 and 7 show the cumulative eroded mass at each altitude during lander descent and the resulting erosion crater profile for three cases of Soil Model 3.

Table 1. Results with Three Soil Models.

| Soil Model | 1 | 2 | | | 3 | | |
|---|---|---|---|---|---|---|---|
| $\rho_0$ (kg/m³) | 1600 | 1100 | | | 1300 | | |
| $\rho_\infty$ (kg/m³) | 1600 | 1800 | | | 1800 | | |
| $H$ (cm) | N/A | 6.8 | | | 8.0 | | |
| Erosion depth (cm) | 515 | 5 | 6 | 12 | 4 | 6 | 12 |
| $\varepsilon$ | 0.0032 | 0.0042 | 0.0042 | 0.0042 | 0.0027 | 0.0035 | 0.0042 |
| $B_1$ | 0.579 | 0.657 | 0.628 | 0.599 | 0.639 | 0.611 | 0.588 |
| $k$ (kg/m³) | N/A | 49.0 | 57.0 | 94.8 | 27.0 | 37.4 | 62.6 |
| $M$ (tonnes) | 195.8 | 11.4 | 13.3 | 22.3 | 10.7 | 15.3 | 25.7 |

Models 2 and 3 produced similar results, showing that the model is not too sensitive to the uncertainties in $\rho_0$ or $H$. $\varepsilon$ was consistently about 0.003 to 0.004 in all cases. $k$ is not constrained well by these results so laboratory measurements and/or careful measurement of erosion depths in lunar landings are needed. The total eroded mass is between 11 and 26 t, which is 4 to 10 times higher than the previous estimate of $M = 2.6$ t [8]. Thus, the damage to surrounding hardware from lunar landing blasts may be worse than previously believed. Additional uncertainties in the analysis are discussed below.

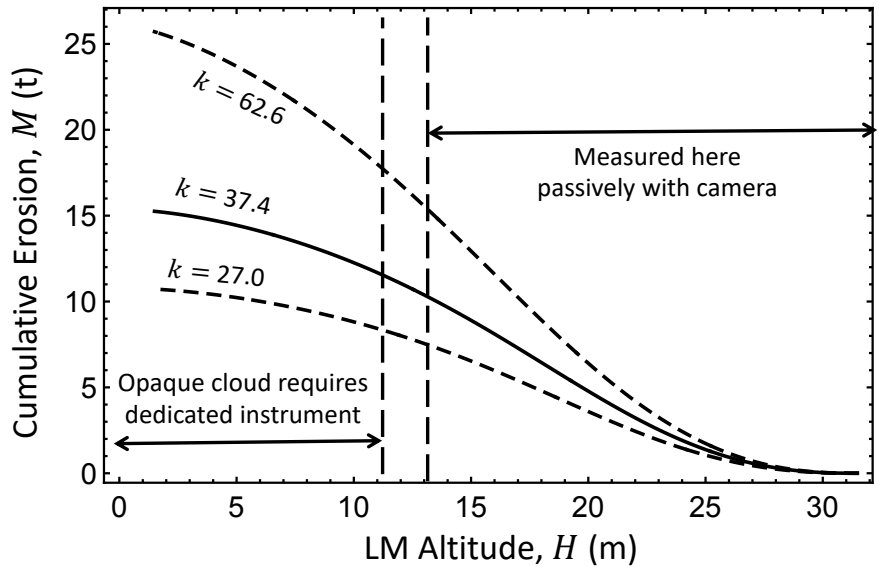

**Figure 6. Cumulative eroded mass during descent versus lander height.** $k$ values in kg/m$^3$.

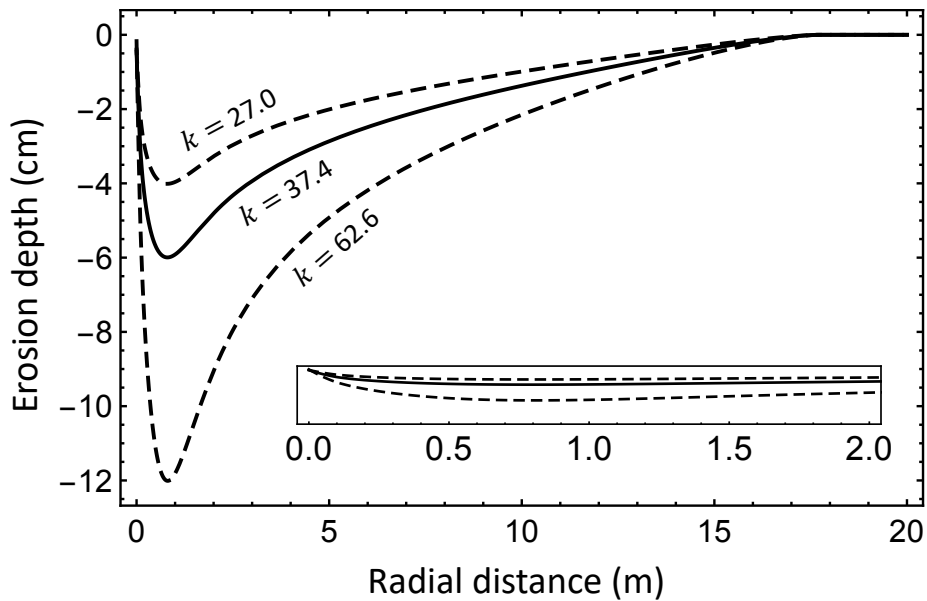

**Figure 7. Erosion crater depth (cm) versus radial distance (m).** Vertical axis greatly exaggerated for clarity. Inset: same plot for the first 2 m radial distance but without vertical exaggeration to show how the crater is gradual and difficult to discern in post-landing imagery. The high point on centerline was probably removed by bulk soil failure in the final moments of landing, leaving it flat across the center. $k$ values in kg/m$^3$.

## 6. Discussion

There are uncertainties in the particle size distribution of the ultra-fines fraction of lunar soil, which has a large effect on optical depth. To estimate the uncertainty, each of the ultra-fine distributions measured by Park et al. (Apollo soils 10084 and 70051 plus JSC-1Avf dust simulant) was substituted in the calculations, varying the mass fraction finer than 10 $\mu$m to values between 3%wt and 12%wt. These two lunar soils are both mature in terms of space weathering and produced only about $\pm 10\%$ error in $M$. Varying the mass fraction of the fines resulted in $+53\%$ to $-27\%$ error in $M$ since the fines dominate both optical density and cohesion. Combined uncertainties, not including the range of possible values for $k$, are $+57\%$ to $-35\%$ in $M$. These results can be improved by better characterization of the ultra-fines in lunar soil, by flying dedicated instruments to measure the blowing dust at lower altitudes with large landers, and by carefully measuring the compaction and cohesion vs. depth in the top 12 cm for soil close to the landing site but not so close that the upper layers were eroded. Due to the stochastic nature of the soil column, multiple such measurements are needed.

This study used a fitting process that mixes unlike fitting parameters. $B_1$ is a camera parameter from lack of calibration. $\varepsilon$ was developed in the companion paper as a constant in erosion physics. $k$ is a parameter dependent on soil particle shapes and granular dynamics affecting packing of the grains. The other soil model parameters ($\rho_0$, $\rho_\infty$, and $H$) are the result of geological history of the site. In the future, with cameras that have calibrated $B_1$ and after the physics parameters $\varepsilon$ and $k$ have been calibrated through laboratory measurements and other methods, the only unknowns will be the landing site geology. The dust brightness curve can then be used to constrain the density curve in the soil column of the landing site, providing insight into the geology. It may be possible to derive more complex and even non-monotonic density curves for the site using this method.

This theory describes surface erosion, not the bulk cratering mechanisms that occur during Mars landings, which might occur on the Moon with larger landers thrusting close to the surface. Those mechanisms include bulk shearing of soil due to bearing capacity failure [37] or diffusion-driven shearing [38], shock impingement fluidization [39], and diffused gas eruption [40]. Bulk cratering did not generally occur in the Apollo landings except possibly in the large blast of soil observed in the final moments of some landings [4], which may have ejected several centimeters at once from close to the centerline. There is still no theory that can predict the onset threshold or rate of the bulk cratering mechanisms.

## 7. Conclusions

The erosion rate equation developed in the companion paper has been partially validated by comparison with optical density of the blowing dust during an Apollo landing. Integrating the erosion rate during vehicle descent to the lunar surface requires a soil model that tells how the soil becomes more resistant to erosion with increased compaction below the surface. Soil models based on thermal inertia of the lunar soil measured by LRO Diviner and on Apollo core tube and drill core data produce excellent fits to the blowing dust, but this requires a parameter that tells how cohesive energy of the soil increases with compaction. That parameter has been crudely constrained here by imagery of the post-landing erosion depths, but the permissible range still varies by a factor of three. This results in a factor of 2.6 uncertainty in the total mass of eroded soil blown during each Apollo landing. This can be improved by measuring $k$ directly in the laboratory. The ultra-fines fraction of lunar soil is also not well-constrained, and this produces another $+53\%$ to $-27\%$ uncertainty in the total eroded soil. The new estimate of eroded soil during Apollo landings is 11 to 26 t, which is 4 to 10 times higher than prior estimates. This suggests that soil blown during lunar landings can cause much worse damage to surrounding hardware than previously recognized.

**Funding**

This work was supported by NASA grant numbers 80NSSC20K0810 and NNA14AB05A.

**Author Contributions**

Conceptualization, formal analysis, writing – original draft, and writing – review and editing: PM

**Competing Interests**

The author declares there are no competing interests.

**Data and Materials Availability**

Tabular data for pixel brightness in fig. 5 are available in the supplementary material. Apollo landing video used in this analysis is available at https://youtu.be/JSXhb3J05ps.

**Supplementary Figure**

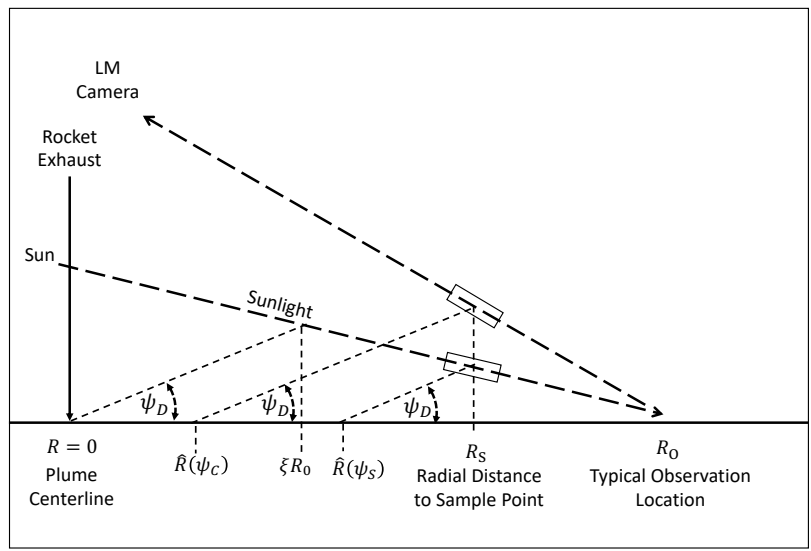

**Supplementary Figure 1**. Three integration limits for eq. 3 determined by the maximum dust angle, $\psi_D$. See text for description.